\documentclass[12pt]{article}

\usepackage{fullpage}

\usepackage{times}
\usepackage{mathptmx}
\usepackage{amssymb}
\usepackage{amsmath}

\makeatletter
\newcommand{\smallbullet}{} 
\DeclareRobustCommand\smallbullet{%
  \mathord{\mathpalette\smallbullet@{0.5}}%
}
\newcommand{\smallbullet@}[2]{%
  \vcenter{\hbox{\scalebox{#2}{$\m@th#1\bullet$}}}%
}
\makeatother

\usepackage{setspace}

\usepackage{graphicx}

\usepackage[table]{xcolor}

\usepackage{flafter}

\usepackage[width=\textwidth]{caption}
    
\usepackage[normalem]{ulem}

\usepackage[small,compact,explicit]{titlesec}

\titleformat{\section} {\vspace{6pt}\sf\bfseries}{\thesection} {12pt} {#1\vspace{-6pt}}
\titleformat{\subsection} {\bf}{\thesubsection} {0pt} {#1\vspace{-6pt}}
\titleformat{\subsubsection}[runin]{\vspace{0pt}\normalfont\normalsize\bfseries}{\thesubsubsection} {6pt} {#1}

\usepackage{enumitem}
\setlist[itemize]{noitemsep, topsep=0pt}
\setlist[enumerate]{noitemsep}

\usepackage{apacite}
\usepackage[sort&compress]{natbib}

\usepackage{url}
\AtBeginDocument{\urlstyle{APACsame}}

\usepackage{tabularx}

\begin{document}

\doublespacing

\begin{flushleft}
{\large\bf Building networks of shared research interests by embedding words into a representation space}

{Art F.~Y.~Poon$^{1,2,3}$}

$^1$Department of Pathology \& Laboratory Medicine, Western University, London, Canada\\
$^2$Department of Microbiology \& Immunology, Western University, London, Canada\\
$^3$Department of Computer Science, Western University, London, Canada\\

\end{flushleft}


\section * {Abstract}

Departments within a university are not only administrative units, but also an effort to gather investigators around common fields of academic study.
A pervasive challenge is connecting members with shared research interests both within and between departments. 
Here I describe a workflow that adapts methods from natural language processing to generate a network connecting $n=79$ members of a university department, or multiple departments within a faculty ($n=278$), based on common topics in their research publications.
After extracting and processing terms from $n=16,901$ abstracts in the PubMed database, the co-occurrence of terms is encoded in a sparse document-term matrix.
Based on the angular distances between the presence-absence vectors for every pair of terms, I use the uniform manifold approximation and projection (UMAP) method to embed the terms into a representational space such that terms that tend to appear in the same documents are closer together.
Each author's corpus defines a probability distribution over terms in this space.
Using the Wasserstein distance to quantify the similarity between these distributions, I generate a distance matrix among authors that can be analyzed and visualized as a graph.
I demonstrate that this nonparametric method produces clusters with distinct themes that are consistent with some academic divisions, while identifying untapped connections among members.
A documented workflow comprising Python and R scripts is available under the MIT license at \url{https://github.com/PoonLab/tragula}.

\clearpage

\section * {Introduction}

Departments within a university faculty are not only administrative units, but also an effort to organize the academic specialties into clusters \citep{dressel1970university}.
A common task for administrators and faculty members is to identify clusters of expertise within a department, or research themes that bridge multiple departments \citep{demes2019catalyzing}.
These clusters provide a useful framework to facilitate collaborations within the university, attract external collaborators and funding \citep{qi2022choosing}, and recruit prospective graduate students and postdoctoral scholars.
Although identifying these clusters is a ubiquitous task, there is a lack of quantitative methods for accomplishing it, and clusters tend to be the result of manual, subjective curation of qualitative data.
Furthermore, a common objective of university administration and faculty groups is to encourage collaborative research between departments, \textit{i.e.}, `breaking down silos' \citep{luke2015breaking}.
A major challenge to establishing these collaborations is identifying individuals in other academic departments who share similar research aims.

There is an extensive literature on identifying and characterizing networks of scientific collaboration through the analysis of publication data, \textit{i.e.}, bibliometrics.
For example, co-authorship networks can be constructed from the sparse matrix of authors and documents, using readily accessible bibliographic databases \citep{fagan2018assessing}.
The drawback to this approach is that it requires that the individuals in the network have already collaborated to the extent that they have co-authored a sufficient number of publications.
If our purpose is to discover overlapping research interests that have yet to be acted upon, then a co-authorship network will likely be of limited utility.
Co-authorship ties also tend to be influenced by shared characteristics other than research interests, including gender and departmental affiliation \citep{fagan2018assessing}.
Consequently, individuals working on similar research themes may be located in disparate parts of the network.
Another cateogry of methods in bibliometrics is author co-citation analysis, which evaluates the frequency that two authors both appear in separate items of a document's bibliography \citep{eom2009introduction}.
However, this approach implicitly assumes that authors are cited at similar rates, which may be violated by differences in authorship and citation norms among fields, self-citation abuse, or the under-representation of certain groups \citep{osareh1996bibliometrics}.


Another popular methodology that has been applied to the problem of identifying clusters from large sparse datasets is topic modeling \citep{yan2012overlaying, yau2014clustering}.
Topic modeling is a statistical method for mapping unstructured data such as text to latent states.
The overall objective is to create a compact representation of the original data without an excessive loss of information.
In many cases this is accomplished with some form of latent Dirichlet allocation \citep[LDA;][]{blei2003latent}, in which we assume there are $k$ topics that each define a probability distribution over the entire vocabulary of words in the corpus.
Each document in turn can be modeled as an unordered collection of words generated by a mixture of these topic distributions.
LDA generally requires the user to specify the number of topics \textit{a priori}, and the challenge of parameterizing the Dirichlet distributions can cause the model to become computationally complex and unstable \citep{vayansky2020review}.
Hierarchical stochastic block modeling is another approach to topic modeling where community detection is used to identify clusters in the bipartite graph induced by the occurrence of terms among documents \citep{gerlach2018network}.
However, our primary objective is to characterize the similarities in research expertise among a set of authors, not the identification of latent topics.




In this study, I describe an efficient non-parametric workflow for inferring clusters of expertise from the publication outputs of members of one or more academic units.
Like many topic modeling approaches, this method generates a document-term matrix to record the co-occurrence of words among documents.
However, rather than attempting to compare documents, or reconstruct topics as latent states that generate documents, I calculate distances between terms based on their co-occurrence in the document-term matrix.
These distances are then used to embed terms into a latent space by uniform manifold approximation and projection (UMAP).
Each author's own corpus thus defines a probability distribution over terms embedded in this space.
This allows us to calculate the Wasserstein (\textit{i.e.}, earth mover's) distance between the distributions for each pair of authors to yield a distance matrix that provides a basis for a standard library of analyses and visualization.
I demonstrate this workflow and visualizations of its outputs by applying it to a corpus of NCBI PubMed abstracts associated with members of an academic department, as well as six other departments within the same university faculty.

\section * {Methods}

\subsection * {Data collection}

I used the BioPython \citep{cock2009biopython} Entrez submodule to query the PubMed database API \citep{coordinators2017database} with the full names of all faculty members of the seven basic science departments within the Schulich School of Medicine and Dentistry, based on the online departmental faculty listings, \textit{e.g.},  \url{https://www.schulich.uwo.ca/pathol/people/faculty/index.html} (last accessed May 27, 2024).
The results of each query were manually screened for evidence of multiple authors, \textit{e.g.}, varying middle initials, inconsistent affiliations.
In these cases, I modified the query to exclude the other authors (\textit{e.g.}, `forename lastname NOT forename A lastname') or to filter by past and current affiliations, \textit{e.g.}, `forename lastname AND (``western university'' OR ``university of waterloo'').
All query strings were recorded in CSV files for automation and reproducibility.
For each author query, I serialized the database ID (PMID), authors, title, abstract and keywords for every publication record retrieved from the database into a JSON (JavaScript Object Notation) file.

Next, I used the Python NLTK \citep[Natural Language ToolKit;][]{bird2009natural} module to process these text data.
For each article, I tokenized the title, abstract and keywords into a list of substrings.
The resulting list was passed through the NLTK part-of-speech tagger that classifies words on the basis of their meaning and context.
All words were converted into lower-case and plural forms were converted into singular forms using the lemmatize function that is supported by the WordNet lexical database \citep{fellbaum2010wordnet}.
Next, I removed any words with an exact match in an exclusion list that I manually curated based on preliminary results (\textit{e.g.}, `what',  `always', `lower'), as well as words that matched a regular expression 
corresponding to numeric values, \textit{e.g.}, `$-0.1$'.
For all words that passed these filters, I recorded the word frequency distributions for every author and serialized the result into a JSON file.

Each document $D_i$ is interpreted as an unordered set of words.
The entire corpus is the collection of documents from all authors, $D=\{D_1, \ldots, D_m\}$, and the vocabulary is produced by pooling all words across documents, $V=\bigcup_i D_i$.
I recorded the overall frequency of every word in $V$ in $D$ and sorted the results in decreasing order, defining a global word index that was exported to a file for later steps.
I used the Nelder-Mead method implemented in the R function \textit{optim} to fit the Zipf-Mandelbrot model to the observed log-log distribution of ordered word frequencies by least-squares minimization.
This model predicts the frequency of the $i$-th most frequent word as $f_i = C (i+\beta)^{-\alpha}$, where $C$ is a normalizing constant \citep{piantadosi2014zipf}.
Finally, I recorded the occurrence of words (terms) in each document as a sparse document-term matrix consisting of three columns for the document index, term index, and term frequency.


\subsection * {Natural language processing}

The natural language processing steps of the analysis were carried out in the R statistical computing environment \citep{rcore2023r}.
To increase computational efficiency, I reduced the vocabulary $V$ to the $n=5,000$ or $10,000$ most frequent words, $V'$.
I used the R package Matrix (\url{https://matrix.r-forge.r-project.org/}) to import the document-term ($m\times n$) co-occurrence matrix $X$ into a compressed sparse object \citep{bates2023matrix}.
All non-zero frequencies were stored as 1, under the assumption that the relevance of a word to matching themes should not be contingent on its repeated use in a given document $D_i$.
I screened for duplicate rows in the co-occurrence matrix caused by the same document appearing multiple times in the corpus, \textit{i.e.}, co-authorship with two or more members of the same department or faculty.
To facilitate screening for duplicates in a large sparse matrix, I calculated the checksum $\sum_{j=1}^{n} x_{ij}j$, where $j$ is equivalent to that term's rank, and $x_{ij}\in\{0,1\}$.
Because terms were sorted in decreasing order of global frequency, larger values of $j$ were associated with increasingly rare terms, preventing this sum from exceeding the maximum integer value of the computing environment.
Rows with identical checksums were then directly evaluated for identity, greatly reducing the number of comparisons.
Duplicate rows were discarded from $X$ to prevent subsequent steps from over-estimating the co-occurrence of terms associated with these repeated entries.

Next, I used the R package \textit{wordspace} \citep{evert2014distributional} to calculate the angular distance $\phi$ between every pair of words in $V'$:
\[
\phi = \cos^{-1} \left( \frac{x^T y}{\sqrt{x^T x} \sqrt{y^T y}} \right)
\]
where $x$ is a binary vector of length $n$ for the presence/absence of word $w_1$ in each document in the corpus $D$, and $y$ is the equivalent for word $w_2$.
The denominator adjusts for the overall prevalence of each word in the corpus.
Note that $\phi$ ranges from 0 to $\pi$ radians.
The result was stored as a compact object of class \textit{dist}.
To generate a reduced dimensional representation of the resulting angular distance matrix, I used the uniform manifold approximation and projection (UMAP) function implemented in the R package \textit{uwot} \citep{melville2024uwot} to embed all words in $V'$ into the real coordinate space $\mathbb{R}^n$.
I evaluated different settings of $n$ between 2 and 6; the results presented here were generated under $n=3$.
UMAP is a non-linear dimensionality reduction algorithm that attempts to maintain the overall shape of the distribution of points as it maps them from high to low dimensional spaces \citep{mcinnes2018umap}.
This algorithm is not only popular in several areas of bioinformatics research \citep[\textit{e.g.},][]{yang2021dimensionality}, but it has also been reported to improve the performance of clustering methods in the context of word embedding \citep{asyaky2021improving}.

Finally, I used the function \textit{wpp} (weighted point pattern) in the R package \textit{transport} \citep{schuhmacher2024transport} to map the word frequency distribution for each author, $f_a(V)$, onto the word embedding produced by UMAP.
A weighted point pattern is akin to a discrete probability distribution over a finite number of points in a continuous space.
To calculate a distance measure between authors, I employed the \textit{wasserstein} function in the \textit{transport} package.
The $p$-Wasserstein (Kantorovich-Rubinstein) distance \citep{panaretos2019statistical} is essentially the smallest amount of work required to transform one probability distribution ($f$) to another ($g$) by transporting probability mass among points, accounting for the distance between points.
Thus, it is also known as the earth mover's distance.
I set $p=2$ so that the cost function for transporting mass takes the Euclidean distance between points; however, similar results were obtained with $p=1$, \textit{i.e.}, the $L_1$-norm.
Since this was the most time-consuming step of the analysis, I employed the R package \textit{parallel} to distribute jobs across multiple cores.

\subsection * {Data visualization}

There are many options for visualizing the matrix $W$ of Wasserstsein distances among authors.
For instance, one can use multidimensional scaling to project $W$ to a two-dimensional scatterplot.
However, I found graph-based visualizations to be more versatile and informative.
I used the R package \textit{igraph} \citep{gabor2025igraph} to construct a graph in which vertices represented authors.
Vertices $i$ and $j$ were connected by an undirected edge if the corresponding distance $W_{ij}$ was below a threshold $\epsilon$.
To ensure that every vertex is connected to at least one other vertex in $G$, I set $\epsilon$ to be the maximum of the shortest distances from each author: 
$\epsilon = \max\left\{\min\{W_{ij}: v_j \in V\backslash v_i\}: v_i \in V\right\}$.
Applying this cutoff induces an adjacency matrix $A$, where
$A_{ij} = 1$ if $W_{ij} < \epsilon$ and $0$ otherwise.
In addition, I simplified the graph by limiting the number of $1$ entries in each row $A_{i\,\smallbullet}$ to the $k$ smallest distances in $W_{i\,\smallbullet}$, \textit{i.e.}, the $k$-nearest neighbours of $v_i$.
Note that this nearest neighbour property is not commutative, \textit{i.e.}, $u$ being a $k$-nearest neighbour of $v$ does not guarantee that $v$ is a $k$-nearest neighbor of $u$.
Finally, I generated a new adjacency matrix where $A'_{ij}=\max\{A_{ij}, A_{ji}\}$.
All Python and R scripts for data collection, natural language processing and visualization are available under the MIT license at \url{https://github.com/PoonLab/tragula}.

\clearpage

\section * {Results}

\begin{figure}[tbp]
\centering
\begin{tabular}{ll}
{\large A} & {\large B}\\[-24pt]
\includegraphics[width=0.45\textwidth]{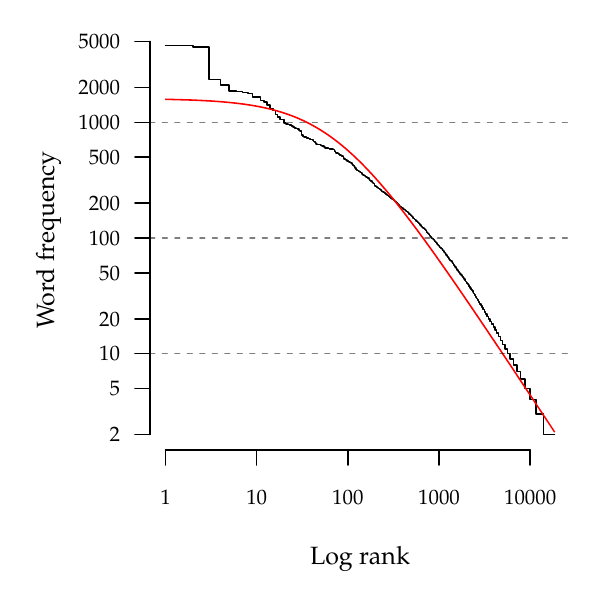} & 
\includegraphics[width=0.45\textwidth]{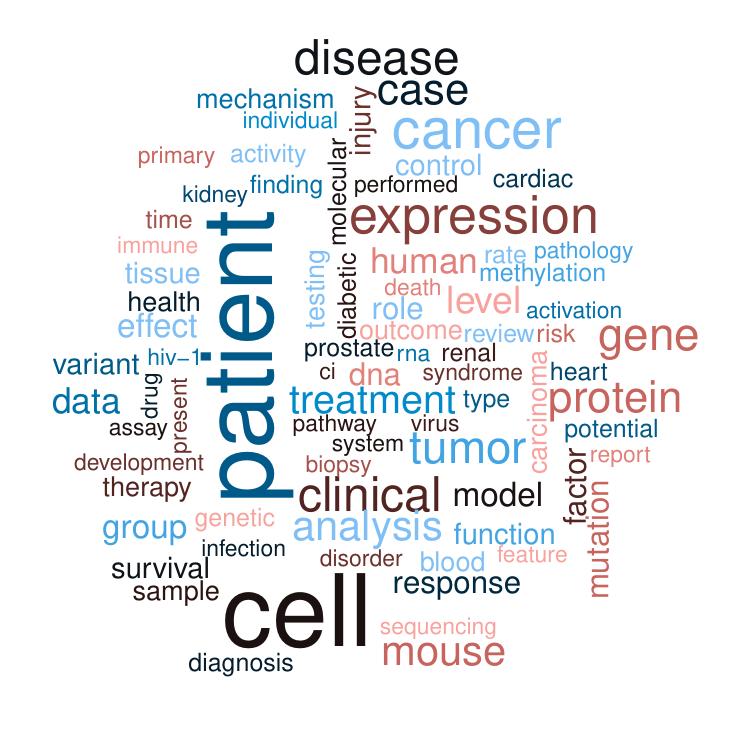}\\
\end{tabular}
\caption{
Word frequency distribution for the Pathology department corpus.
(A) A log-log plot of word frequency against rank for words sorted in decreasing order of frequency.
The red curve represents a least-squares fit of a Zipf-Mandelbrot model ($\alpha=1.2$, $\beta=72.7$, $C=2.73\times 10^{5}$) to these log-transformed values.
(B) A wordcloud depicting the 80 most frequent words in this corpus.
}
\label{fig:worddist}
\end{figure}

\subsection * {Department-level analysis}

I collected a total of 16,901 documents derived from the PubMed abstracts, titles and keywords for the publications associated with 282 faculty members of the seven departments.
The distribution of faculty members and abstract data among these departments is summarized in Supplementary Table \ref{tab:faculty}.
First, I carried out an analysis of the corpus for members of the Pathology department only.
A document-term co-occurrence matrix was generated for 18,478 words passing the filtering criteria and appearing more than once.
As typical, the frequency distribution of words in the entire corpus was highly skewed towards low numbers (Figure \ref{fig:worddist}A).
For example, 12,434 (67\%) of words appeared fewer than 10 times across 3,530 documents.
`Patient' and `cell' were the two most commonly used words for this department.
Among documents, these two words were significantly negatively associated (Fisher's exact test, odds ratio $= 0.69$, 95\% confidence interval $0.59 - 0.80$).
In other words, a scientific abstract mentioning cells was significantly less likely to also refer to patients.
Conversely, an abstract mentioning cancer was significantly more likely to refer to patients ($\text{OR}=2.54$, 95\% CI $=2.13-3.02$).

\begin{figure}[tbp]
\centering
\includegraphics[width=\textwidth]{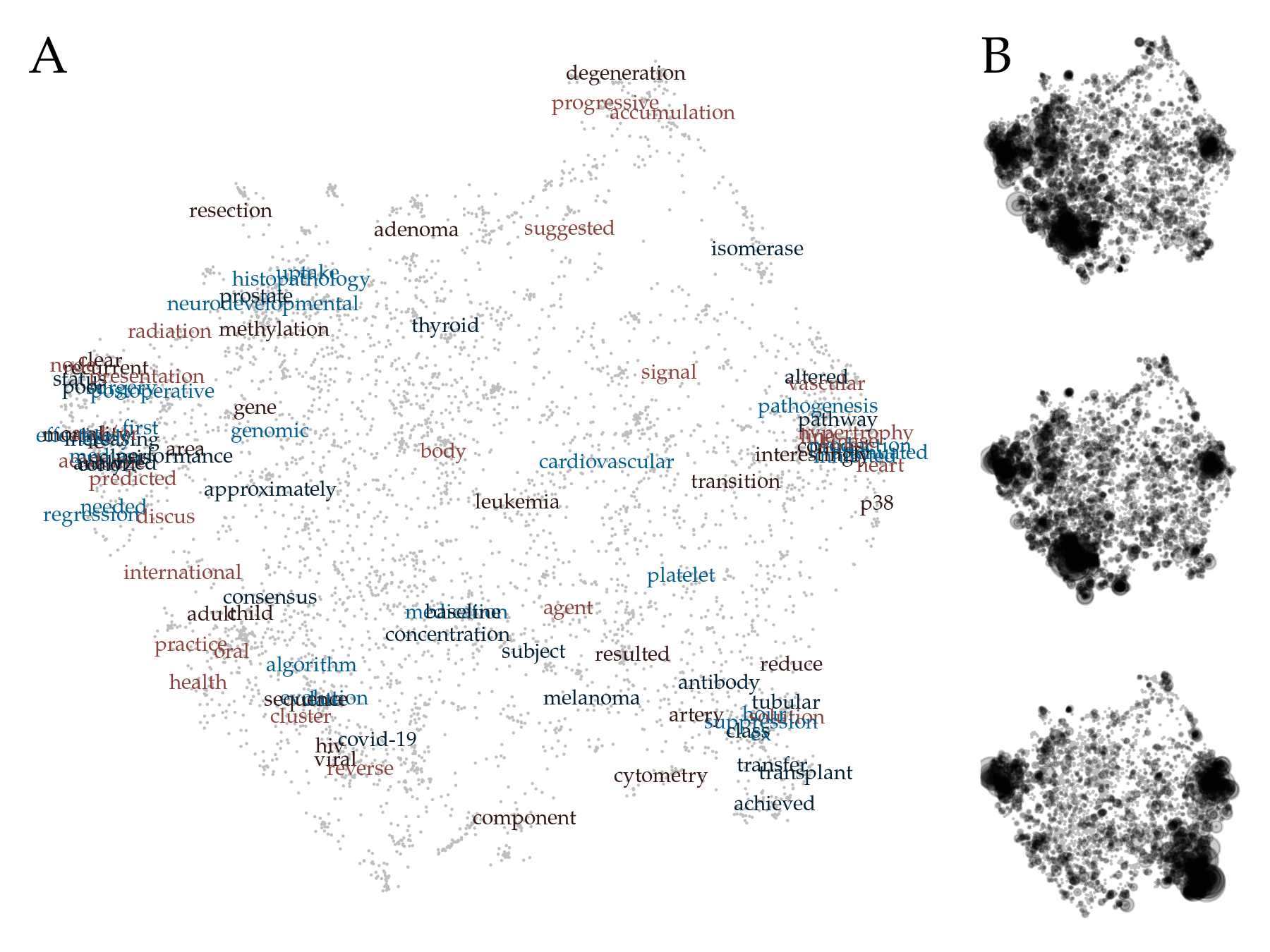}
\caption{
Uniform manifold approximation and projection (UMAP) of the term-term angular distance matrix for the Pathology department.
(A) A random selection of 100 out of the 1,000 most frequent terms are labeled with the respective words, identifying some topical clusters.
These are coloured at random to help visually differentiate overlapping words.
(B) `Fingerprints' of three different members of the Pathology department, produced by rescale the areas of the UMAP points in proportion to each member's word frequency distribution.
The top and middle members were selected for their \textit{a priori} similar research interests, with a third member (bottom) chosen arbitrarily from the other departmental members. 
}
\label{fig:topicspace}
\end{figure}

I discarded 430 (12.2\%) rows from the co-occurrence matrix corresponding to duplicate documents due to co-authorships among departmental members.
Next, I derived a term-term angular distance matrix from these filtered co-occurrence data.
Since low-frequency words would be less informative for comparing documents, I limited the vocabulary to the 5,000 most frequent words (representing 86\% of the corpus) to reduce computing time.
This distance matrix induces a high-dimensional space in which words that tend to appear in the same documents occupy similar locations.
Figure \ref{fig:topicspace}A depicts the uniform manifold approximation and projection of this matrix, with a random selection of points labeled by the respective words.
These labels reveal clusters of related words, such as \{`hiv-1', `covid', `viral'\} or \{`degeneration', `progressive', `accumulation'\}, corresponding to topics in this representation space.

Finally, I calculated the Wasserstein (earth mover's) distance between every pair of members.
Each member's word frequency distribution can be rescaled to a discrete probability distribution over a set of points embedded in the representation space.
Figure \ref{fig:topicspace}B depicts these probability distributions for three departmental members as individual `fingerprints'.
The first two members were selected because their research interests were known \textit{a priori} to be highly similar, while a third was selected arbitrarily.
The Wasserstein distance represents the smallest amount of work required to transform one person's word frequency distribution into another.
Less work is required to shift probability mass between nearby points.
Consequently, members are not penalized severely for using different words if those words are related to a common topic.
In other words, we are not simply comparing frequency distributions on a word-by-word basis.

\begin{figure}[tbp]
\centering
\begin{tabular}{ll}
{\large A} & {\large B}\\[-12pt]
\includegraphics[width=.55\textwidth]{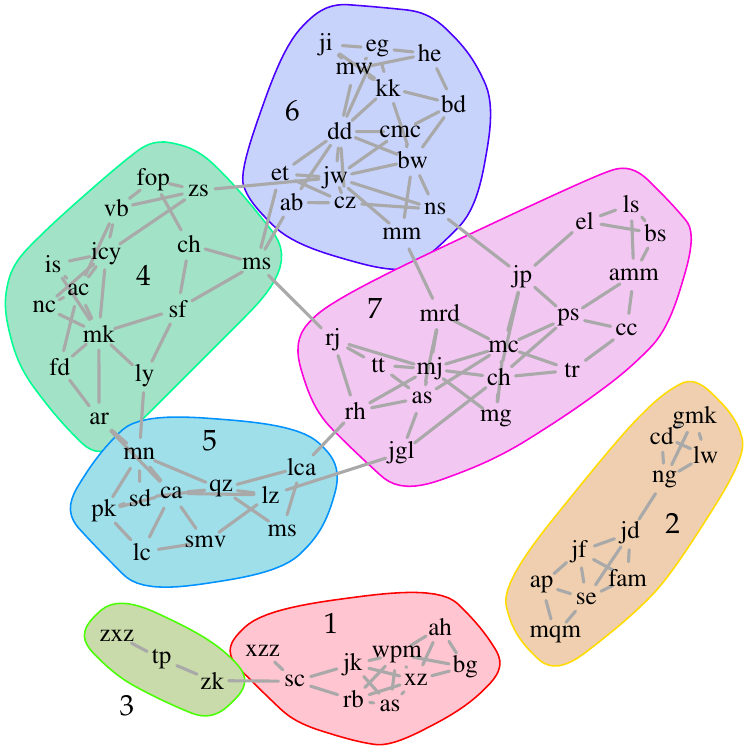} & 
\includegraphics[width=.42\textwidth]{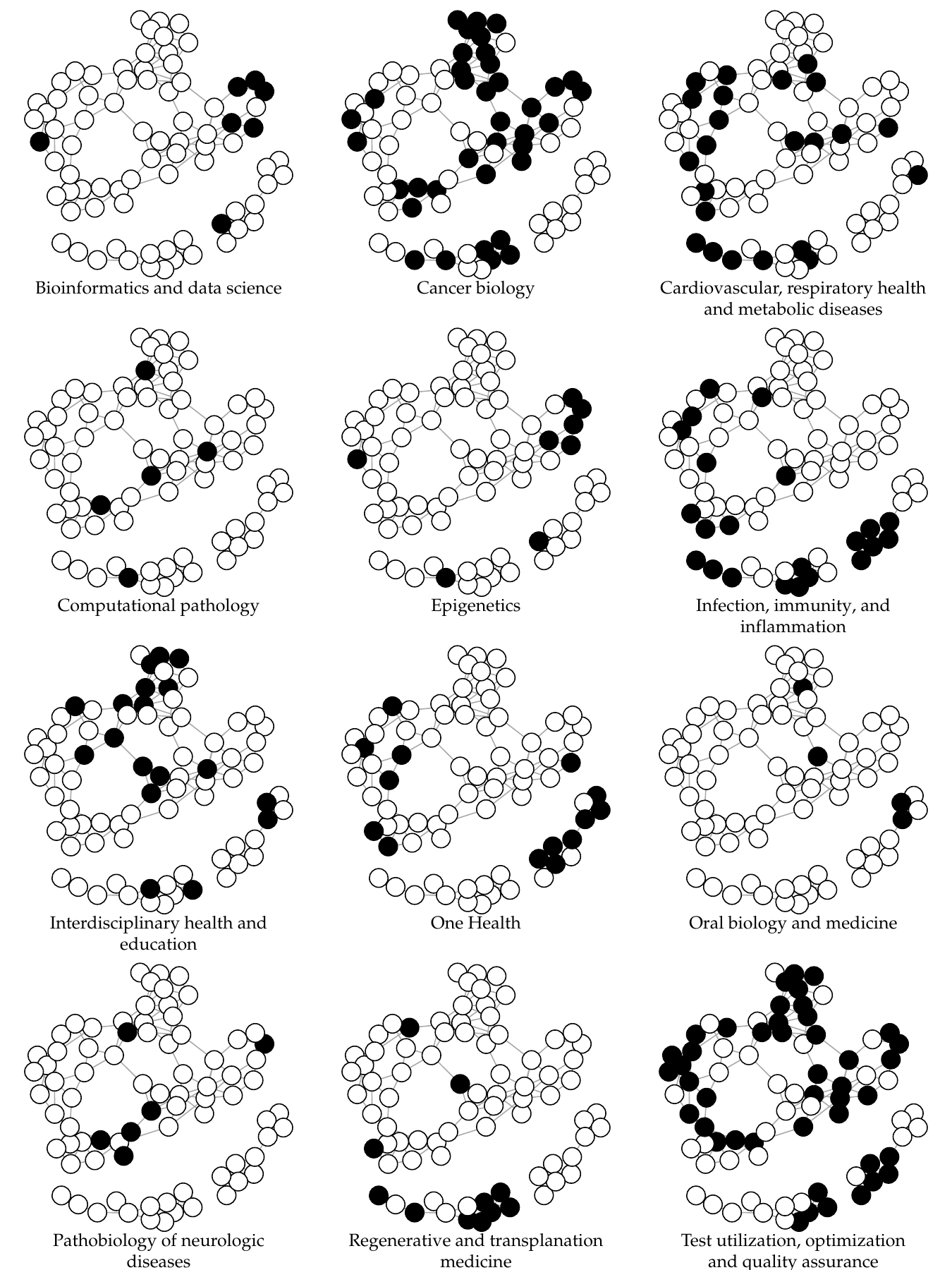}\\
\end{tabular}
\caption{
(A) Undirected graph depicting $n=79$ members of the Pathology department.
Each departmental member is represented by a vertex (labeled with that member's initials). 
Outgoing edges connects each vertex to the three other vertices with the shortest Wasserstein distances.
Coloured regions represent clusters (1-7) produced by the Louvain community detection method.
This layout was generated using the default Fruchterman-Reingold algorithm in igraph.
(B) Reproductions of the same graph with vertices labeled with each of 12 research areas for the department that were previously manually curated from questionnaire data.
}
\label{fig:knn}
\end{figure}

\subsection * {Graph visualization}

Figure \ref{fig:knn}A displays a graph where up to $k=3$ edges are drawn from each member (vertex) to other members when the corresponding Wasserstein distance is below a threshold $\epsilon$.
This threshold was set to the minimum distance at which every vertex was connected to at least one other vertex in the graph.
Otherwise, any isolated vertices would preclude relating those members to  any of their peers.
For the Pathology department dataset, this requirement was attained at a threshold of $\epsilon=0.94$.
Because the edge density at this threshold was high, we limited each member (ego) to a maximum of three nearest neighbours (\textit{i.e.}, vertices with the shortest Wasserstein distances to the ego) to make the graph easier to visualize and resolve clusters.
Finally, we employed the Louvain method to cluster vertices based on the localized density of edges (modularity).
This method tended to yield seven clusters (Figure \ref{fig:knn}A), although 51 out of 1,000 randomly seeded replicates yielded an eighth cluster comprising four members of cluster 4 (`ch', `ms', `rj' and `sf').
Table \ref{tab:clusters} characterizes these clusters with respect to the ten words that are the most `characteristic' of each cluster, \textit{i.e.}, words that have a markedly higher relative frequency in the cluster than the overall corpus.

\begin{table}[tbp]
\rowcolors{2}{gray!15}{white}
\centering
\begin{tabular}{lrp{3.6in}p{1.2in}}
Cluster & Size & Top 10 characteristic words & Classification\\
\hline
1 & 9 & pin1, dc, sirna, rejection, et-1, graft, dendritic, silencing, allograft, recipient & Transplantation, gene therapy\\
2 & 10 & hiv, hiv-1, sars-cov-2, cluster, virus, viral, strain, resistance, sequence, infection & Virology, microbiology\\
3 & 3 & calpain, cardiomyocytes, myocardial, macrophage, cardiac, dysfunction, heart, apoptosis, prevented, fn & Cardiovascular pathology\\
4 & 14 & iop, glaucoma, transfusion, intervention, metabolic, concentration, cost, child, test, improvement & Opthalmic pathology\\
5 & 10 & amyotrophic, lateral, al(s), spinal, neuron, motor, sclerosis, tau, neuronal, cognitive & Neuropathology\\
6 & 15 & cystectomy, endometrial, radical, bladder, lymph, node, prostate, biopsy, recurrence, chemotherapy & Cancer\\
7 & 18 & methylation, copy, signature, genomic, dna, epigenetic, pathogenic, variant, disorder, ng(s) & Genomics, molecular pathology\\
\hline
\end{tabular}
\caption{
Characterization of numbered clusters extracted by Louvain method from the graph depicted in Figure \ref{fig:knn}.
`Size' provides the number of departmental members in each cluster.
Characteristic words were determined by comparing the word frequency distribution for all members of the $i$-th cluster, $f_i$ to the global frequency distribution $f$ in the form of a relative error, $(f_i-f)/f$, and then sorting all words with respect to this quantity in descreasing order.
The third column of this table presents the ten words with the highest relative error for each cluster.
In two cases, an `-s' suffix was incorrectly removed as a pluralization by the lemmatizer.
`Classification' provides subjective descriptors of predominant themes for each cluster, based on the characteristic words.
}
\label{tab:clusters}
\end{table}

For comparison, we highlighted vertices in the graph for each of 12 research areas that were manually curated by the departmental administration from self-administered questionnaires that were submitted by members of the department (Figure \ref{fig:knn}B).
Some of the more specialized research areas tended to map to distinct subgraphs, such as `bioinformatics and data science' and `regenerative and transplantation medicine', whereas others were more diffuse, \textit{e.g.}, `computational pathology'.
Note that these research areas were subjective and some members were clearly misclassified, so they should not be interpreted as `ground truth'.


Although it is certainly possible to generate angular distances between authors by the direct comparison of their word frequency distributions, that approach risks placing too much emphasis on the exact word choice of authors, instead of their common topics of research.
Put another way, a direct comparison would separate an author with a high frequency of `sequence' but a low frequency of `evolution' from a second author with the opposite word usage, even though these words are topically closely related (Figure \ref{fig:topicspace}A).
For comparison, I evaluated results from direct word-to-word angular distances between departmental members using the same workflow that I employed for Wasserstein distances.
While the direct distances are significantly correlated with the Wasserstein distances (Spearman's $\rho=0.396$, 95\% CI $=0.366-0.426$), this correlation is disproportionately driven by a small number of short word-to-word distances, and there is generally much less variation among word-to-word distances for a given author (Supplementary Figure \ref{fig:word2word}A).
In comparison to the graph depicted in Figure \ref{fig:knn}A, applying the same distance threshold and nearest-neighbour criteria to these direct word-to-word distances yielded a graph with more `hub' vertices (Supplementary Figure \ref{fig:word2word}B).
The degree size distribution of this new graph had a longer tail (range $1-18$, mean $4.7$) than the original graph (range $1-9$, mean $4.0$).
Moreover, the closeness centrality (reciprocal of the sum of shortest path lengths to every other vertex) was significantly higher for vertices in the word-to-word graph than the original (paired Wilcoxon test, $P=6.64\times 10^{-6}$), excluding vertices associated with smaller components in the latter.


\begin{figure}[tbp]
\centering
\includegraphics[width=0.9\textwidth]{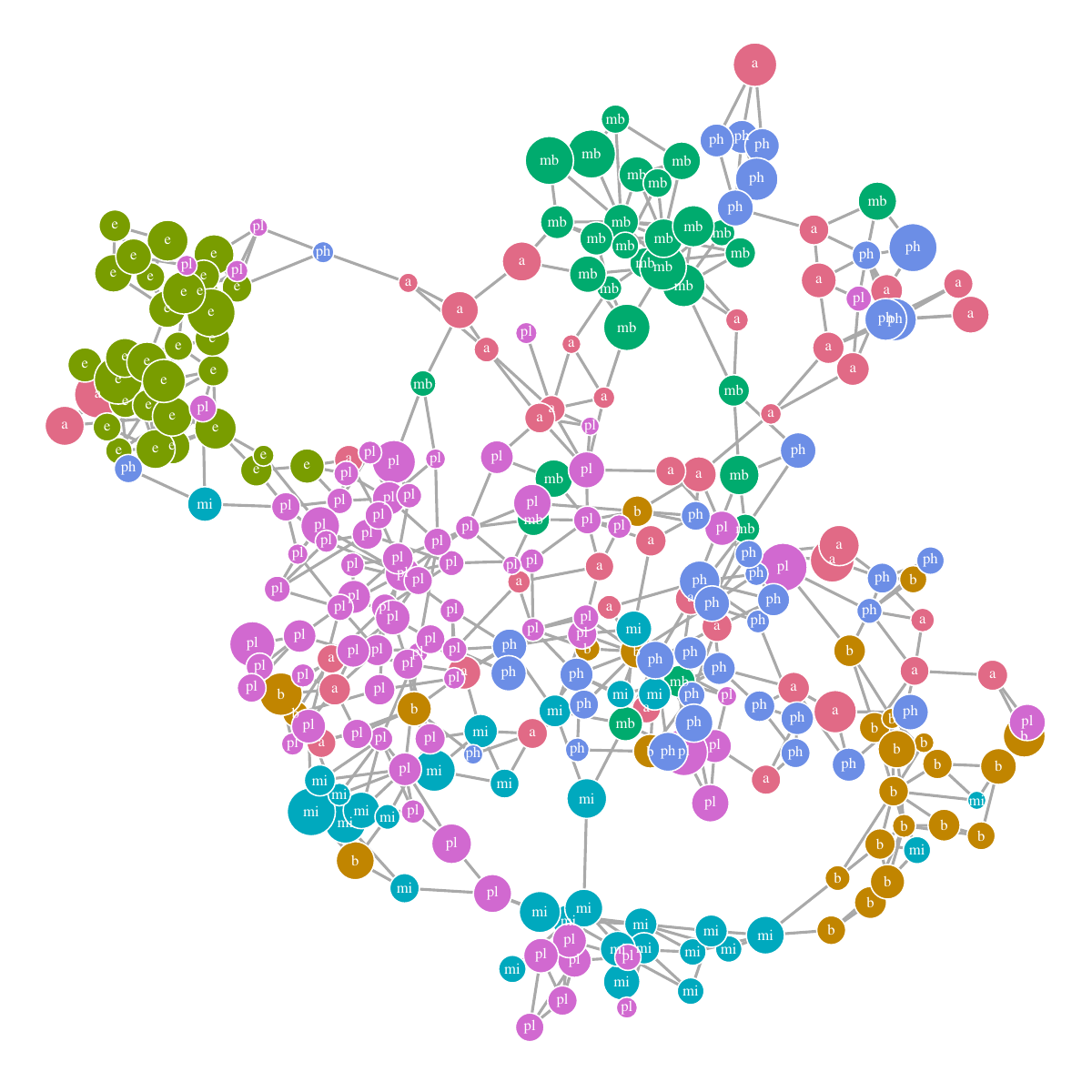}
\caption{
Undirected graph of $n=278$ faculty members in seven basic science departments of the Faculty of Medicine \& Dentistry.
This visualization uses the \cite{kamada1989algorithm} layout algorithm, and limits each vertex to its $k=3$ nearest neighbours.
Each vertex represents a member, with colour and label indicating their primary departmental affiliation: `a' = Anatomy and Cell Biology, `b' = Biochemistry, `e' = Epidemiology and Biostatistics, `mb' = Medical Biophysics, `mi' = Microbiology \& Immunology, `pl' = Pathology \& Laboratory Medicine, `ph' = Physiology \& Pharmacology.
Vertex area is scaled in proportion to the number of abstracts retrieved for each member.
}
\label{fig:faculty}
\end{figure}

\subsection * {Faculty-level analysis}
Lastly, I expanded this analytical workflow to all $n=278$ members of the seven basic science departments in the Faculty of Medicine \& Dentistry.
Because extending the corpus to these additional departments would greatly expand the vocabulary, I increased the word limit from 5,000 to the 10,000 most frequent terms (Supplementary Figure \ref{fig:faculty-wordcloud}).
Figure \ref{fig:faculty} displays the graph induced by applying a distance threshold of $\epsilon=0.938$ to the Wasserstein distance matrix, limiting each row to the three smallest non-zero entries in the resulting asymmetric adjacency matrix, and then converting the directed graph represented by that matrix into an undirected graph.
Some basic science departments formed distinct clusters in this graph.
For example, the subgraph for the Epidemiology and Biostatistics department is both more isolated from the rest of the graph, and has a high edge density among members (Figure \ref{fig:faculty}).
The closeness centrality, which measures how close a vertex is to every other vertex in the graph, was markedly lower for Epidemiology and Biostatistics (mean $0.208$) than the other departments (range $0.222-0.262$, Supplementary Figure \ref{fig:graphstats}).

Similarly, the mean within-cluster distance, \textit{i.e.}, the unweighted path length (number of edges) between members, was lower for this department ($2.88$) than the others (range $3.99-6.51$).
In contrast, both the Anatomy \& Cell Biology and Pathology \& Laboratory Medicine departments were distributed throughout the graph, reflected by relatively high closeness centrality and within-cluster distance statistics.
Nevertheless, all departments exhibited much shorter within-cluster distances on average than expected from randomized null distributions.
The Louvain method for community detection partitioned the undirected graph into 14 clusters (Supplementary Figure \ref{fig:faculty-clusters}).
Generally, the concordance between departments and these clusters was quite low (adjusted Rand Index $=0.20$).
There were, however, distinct concentrations of departmental memberships in specific clusters (Supplementary Table \ref{tab:faculty-clusters}).
For example, 17 of 29 (59\%) of members affiliated with  Medical Biophysics were placed into one cluster.
An interesting result of this particular partition is that the Epidemiology and Biostatistics department was split into two clusters.
Nevertheless these clusters appeared to be very similar, sharing 16 out of the top 30 characteristic words.
Characteristic words that were not shared included `risk', `mortality' and `interval' in cluster 5, and `family', `community' and `medical' in cluster 12.

\section * {Discussion}

The workflow described in this study focused on the use of existing resources (\textit{e.g.}, APIs, packages in Python and R) to enable the efficient computation of distance matrices for a set of authors using conventional desktop hardware.
The most time-consuming computational step in this workflow is the calculation of Wasserstein distances.
For example, it took 7.2 minutes on an AMD Ryzen 9 7950X processor to calculate this distance for $n=79$ authors, given a vocabulary limited to the 5,000 most frequent words.
This was reduced to 1.1 minutes by distributing the tasks across 24 threads, but computing the matrix in parallel also greatly increases the memory requirement of the task.
Computing this distance matrix for $n=278$ authors and an expanded vocabulary of 10,000 words required 31.6 minutes on 24 threads, and roughly 32GB of RAM.
The time required to populate this pairwise distance matrix unavoidably increases quadratically with the number of authors.
In addition, calculating the Wasserstein distance is reported to have cubic time-complexity with the number of points \citep{bernton2019approximate}, although I found that computing time increased roughly linearly with vocabulary size in practice.
While faster approximations to the Wasserstein distance are available \citep[\textit{e.g.}, Sinkhorn distance;][]{cuturi2013sinkhorn}, experiments with an implementation of the Sinkhorn distance in R provided a substantial speed-up only at the cost of grossly overestimating distances and substantial memory consumption.


In fact, the most time-consuming part of workflow was not automated: manually composing PubMed database queries for each author required several days of work.
A query should retrieve only publications associated with a specific person.
However, many people with records in the PubMed database can share the same name, and the same person may publish under different names over their career.
In some cases, members published under a different first or last name than the name associated with their faculty profiles.
This is a long-standing problem in bibliometrics \citep{ferreira2012brief}.
While there are efforts to promote the adoption of unique identifiers such as ORCID \citep{mering2017correctly}, not all researchers have obtained an identifier or enabled public access to their publication list in the associated database.
In the case of this study, the list of publications retrieved by an initial query using an author's first and last names was manually reviewed for consistency.
If there were publications that were not consistent with belonging to that author (\textit{e.g.}, different middle initials, affiliation or subject matter incompatible with their departmental biography), then I used AND or NOT operators to modify the query until the resulting publications were consistent.
This trial-and-error process is difficult to automate.
Ideally, each member would provide their own PubMed query string, or a list of publications in a structured, convertible format such as BibTeX or RIS \citep{hull2008defrosting}. 


A limitation of the analysis presented here is that the corpus was restricted to the title, keywords and abstract of research articles.
The full texts of articles would provide a more comprehensive representation of each author's research interests and expertise.
However, these texts were not consistently available in the NCBI PubMed database due to access restrictions imposed by some journals.
A potential drawback is that abstracts may lack methodological details that are characteristic of a given author, such as the use of a specific experimental technique or class of statistical model.
In addition, a substantial number of PubMed records contained empty fields for abstracts (average 8.6\%, range 0\% to 64.6\% in Pathology).
These records tended to correspond to editorials, short letters and commentaries, rather than contributions reporting findings from primary research.
Another limitation is that the analysis was restricted to works that were indexed in the PubMed database.
Some authors may have contributed articles to journals that are not indexed because they focus on disciplines outside of the biomedical or life sciences, \textit{e.g.}, applied mathematics, computer science, or sociology.
For example, one member of the Pathology and Laboratory Medicine department had several contributions to journals in psychology and international policy that were not indexed in PubMed.


In this study, I have presented clustering results derived from the Louvain community detection method.
There are many community detection methods that can be applied to an undirected graph.
We have no reason to believe \textit{a priori} that the Louvain method is a better choice than any other method, especially in the absence of any ground truth.
One of the limitations of the Louvain method is that it can only produce non-overlapping clusters.
Community detection methods that can accommodate overlapping clusters, where a vertex can be a member of two or more clusters, may provide a more accurate representation for authors who contribute to multiple areas of research.
Nevertheless, it is not the objective of this study to determine which community detection method is the most effective.
Instead, our objective was to generate an informative distance matrix of authors based on the topical similarity of their publications.
Once we have obtained this matrix, there is an enormous library of analysis and visualization tools that can be applied to these distances.
Like most unsupervised clustering problems, these tools require the user to make a number of subjective decisions, \textit{e.g.}, the threshold to convert the distances into an adjacency matrix.
However, these decisions are a part of exploratory data analysis --- arguably more stimulating, and certainly less time-consuming, than the conventional process of manually reviewing the publications and biographies of faculty members to subjectively identify clusters of expertise.







\clearpage

\bibliography{main}
\bibliographystyle{apacite}

\clearpage 
\section * {Supplementary Figures}

\renewcommand{\thefigure}{S\arabic{figure}}
\setcounter{figure}{0}

\begin{figure}[htbp]
\centering
\begin{tabular}{ll}
{\large A} & {\large B} \\[-24pt]
\includegraphics[width=.5\textwidth]{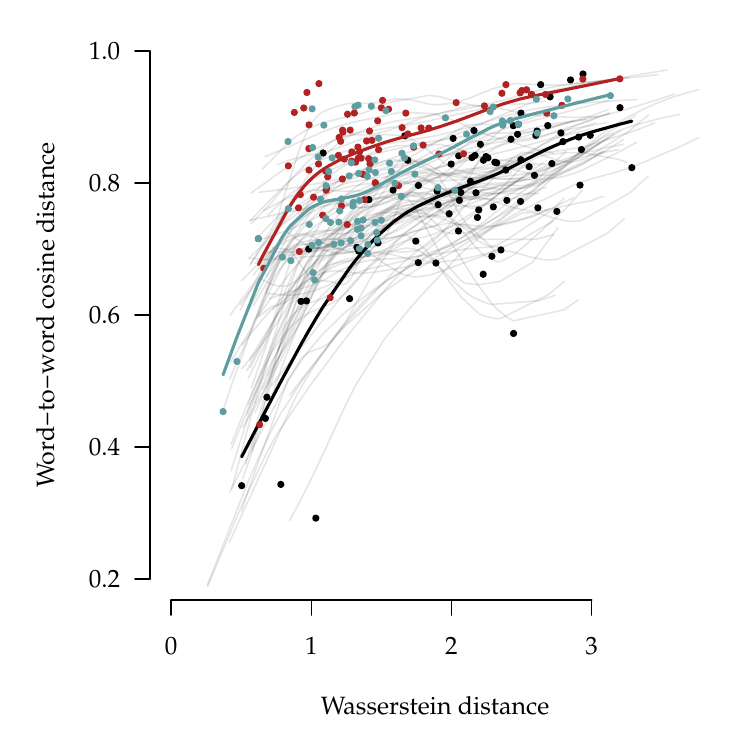} & 
\includegraphics[width=.45\textwidth]{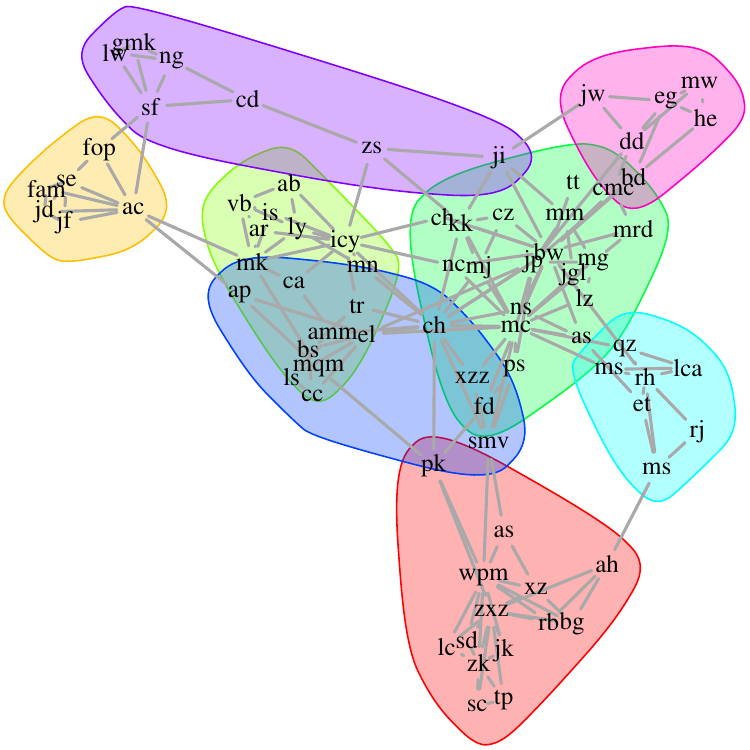}\\
\end{tabular}
\caption{
(A) Scatterplot comparing the word-to-word angular distances against Wasserstein distances for word frequency distributions informed by document-term co-occurrence.
Points are displayed for the first three members, with trendlines generated by smooth splines.
Trendlines for each of the remaining members are displayed in faint grey.
(B) Undirected graph depicting $n=79$ members of the Pathology department.
Each member is represented by a vertex (labeled with that member's initials). 
Outgoing edges connects each vertex to the three other vertices with the shortest angular distances by a direct comparison of author word frequency distributions.
Coloured regions represent clusters produced by the Louvain community detection method.
This layout was generated using the default Fruchterman-Reingold algorithm in \textit{igraph} (version 2.1.3, \citet{gabor2025igraph}).
}
\label{fig:word2word}
\end{figure}

\begin{figure}[htbp]
\centering
\includegraphics[width=5in]{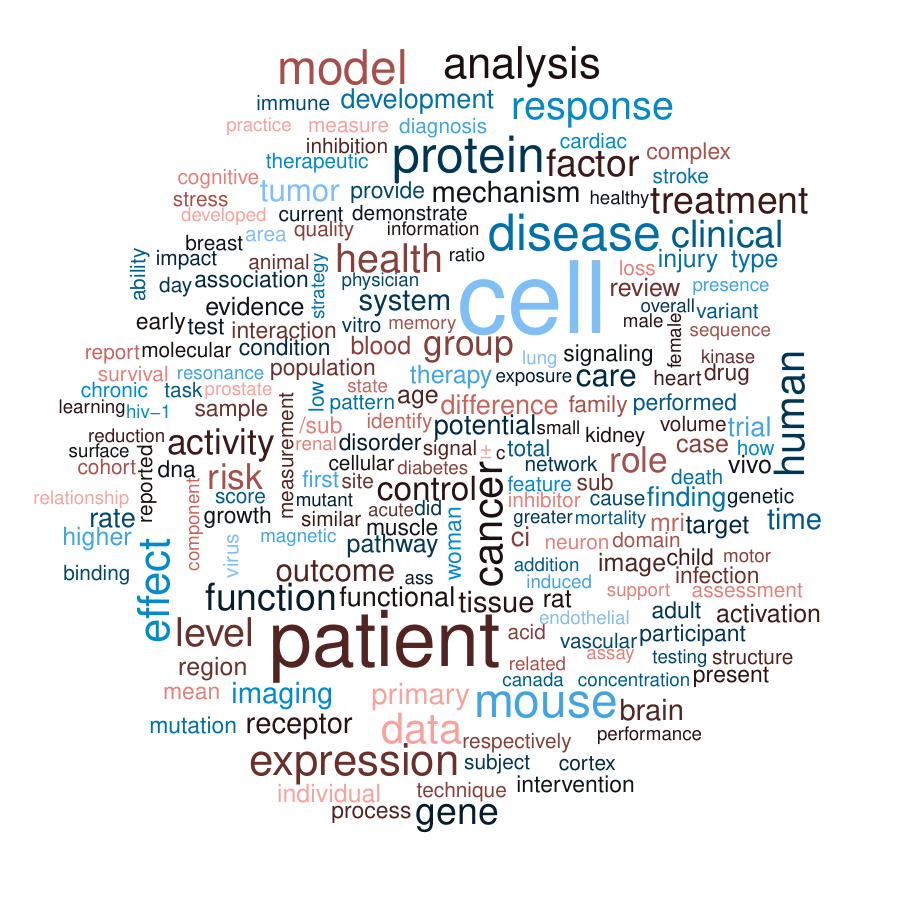}
\caption{
A wordcloud of the 200 most frequent words in the corpus of all $n=278$ members of the seven basic science departments in the Faculty of Medicine \& Dentistry.
Each word is scaled in proportion to its frequency.
This visualization was constructed using the R package \textit{wordcloud} (version 2.6, I.~Fellows, \protect\url{https://cran.r-project.org/package=wordcloud}).
}
\label{fig:faculty-wordcloud}
\end{figure}

\begin{figure}[htbp]
\centering
\includegraphics[width=6in]{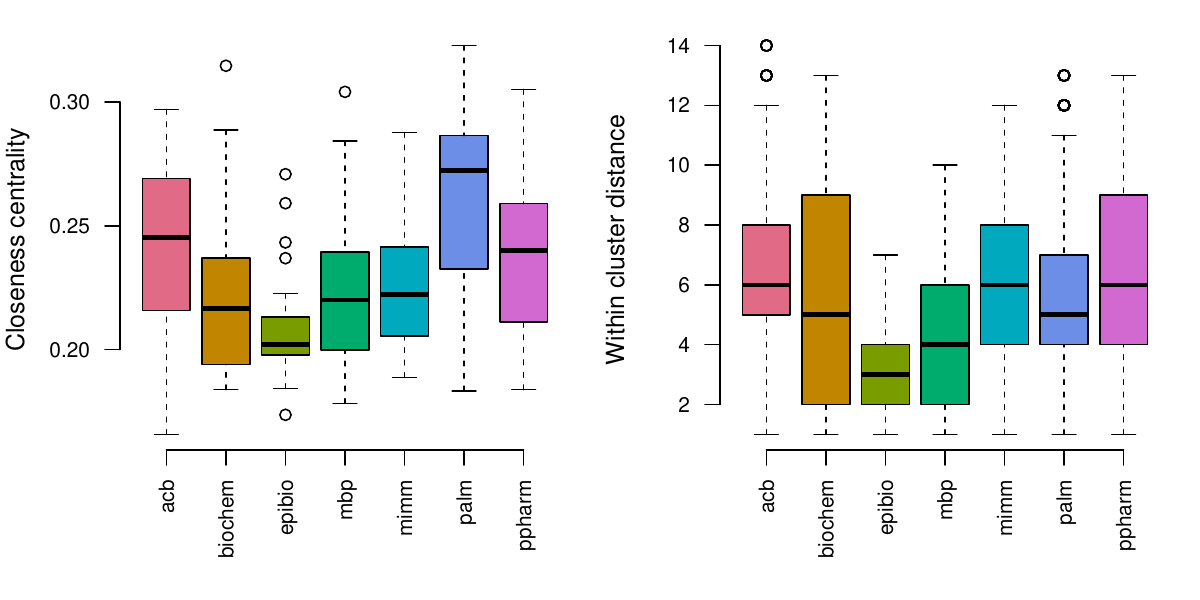}
\caption{
Box-and-whisker plots of closeness centrality (the reciprocal of the sum of path lengths to all other vertices in the graph) and within-cluster distance (the number of edges from a vertex to every other vertex in the cluster) for the seven basic science departments.
}
\label{fig:graphstats}
\end{figure}

\begin{figure}[htbp]
\centering
\includegraphics[width=5in]{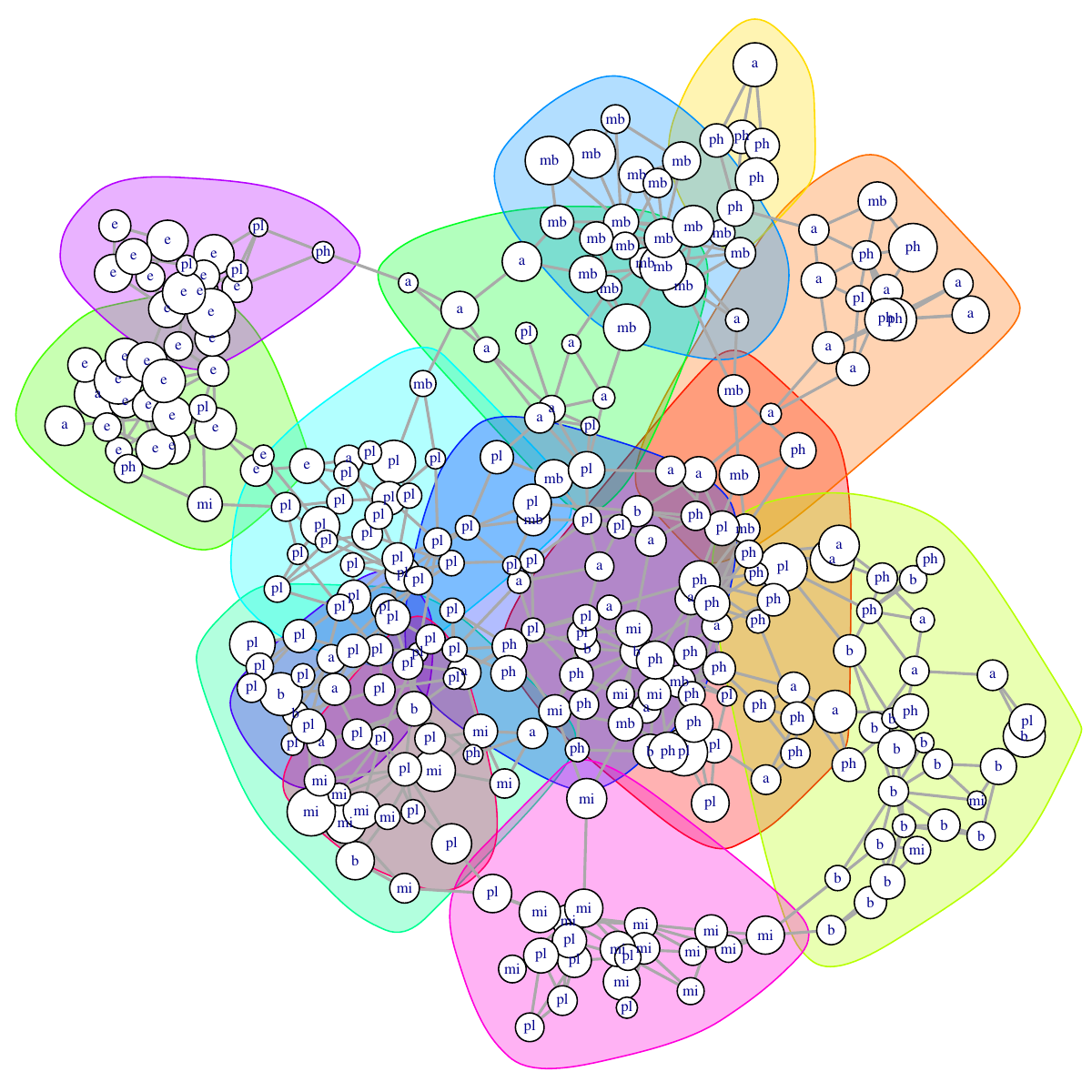}
\caption{
An identical layout of the undirected graph of $n=278$ faculty members depicted in Figure \ref{fig:faculty}, with coloured convex hulls indicating the assignment of vertices to clusters obtained by the Louvain community detection method.
Although the clusters appear to overlap in this diagram, the Louvain method does not produce overlapping clusters (where a vertex is a member of two or more communities).
Thus, the apparent overlaps are due to the limitations of visualizing the graph in two dimensions.
A breakdown of the composition and characteristic words for each cluster is provided in Supplementary Table \ref{tab:faculty-clusters}.
}
\label{fig:faculty-clusters}
\end{figure}

\clearpage 
\section * {Supplementary Tables}

\renewcommand{\thetable}{S\arabic{table}}
\setcounter{table}{0}



\begin{table}[hbtp]
\centering
\begin{tabular}{lrrr}
Department & Members & Abstracts & Median per member\\
\hline
Anatomy \& Cell Biology & 43 & 2,419 & 44.0\\
Biochemistry & 26 & 1,385 & 46.5\\
Epidemiology \& Biostatistics & 31 & 2,620 & 79.0\\
Medical Biophysics & 29 & 2,429 & 69.0\\
Microbiology \& Immunology & 31 & 2,019 & 57.0\\
Pathology \& Laboratory Medicine & 79 & 3,530 & 32.0\\
Physiology \& Pharmacology & 39 & 2,494 & 59.0\\
\hline
\end{tabular}
\caption{
Basic statistics for the number of faculty members and abstracts by department.
These data were retrieved from the NCBI PubMed database on June 6, 2024, with the exception of the Department of Pathology and Laboratory Medicine, for which data were retrieved on May 27, 2024.
Note that the Pathology department combines both basic science and clinical faculty members.
}
\label{tab:faculty}
\end{table}

\begin{table}[hbtp]
\rowcolors{2}{gray!15}{white}
\centering
\begin{tabular}{lrrrrrrrp{3in}}
Cl. & a & b & e & mb & mi & ph & pl & Top five characteristic words\\
\hline
1 & 5 & 3 & & 4 & 1 & 17 & 6 & differentiation, muscle, signaling, stem, metabolic\\
2 & 10 &  &  & 1 &  & 4 & 2 & cognitive, rat, brain, state, neuronal\\
3 & 1 &  &  &  &  &  5 &  & muscle, motor, neuron, area, network\\
4 & 7 & 18 &  &  & 2 & 6 & 2 & domain, binding, interaction, phosphorylation, pin1\\
5 & 2 &  & 18 &  & 1 & 1 & 1 & intervention, trial, prevalence, child, index\\
6 & 8 &  &  & 3 &  &  & 3 & mri, image, brain, imaging, cognitive\\
7 & 3 & 2 &  &  & 7 &  1 & 8 & breast, strain, transcription, line, healthy\\
8 & 1 &  & 1 & 1 &  &  & 24 & endometrial, recurrence, iop, glaucoma, transfusion\\
9 & 1 &  &  & 17 &  &  &  & image, mri, imaging, mm, $\pm$\\
10 & 3 & 1 &  & 3 & 3 & 4 & 10 & mri, imaging, image, spinal, brain\\
11 & 2 & 2 &  &  &  &  &  7 & breast, polymorphism, network, woman, lung\\
12 &  &  & 12 &  &  & 1 & 3 & care, child, intervention, family, health\\
13 &  &  &  &  &  14 &  & 8 & hiv-1, cd4, donor, virus, graft\\
14 &  &  &  &  &  3 &  & 5 & cluster, hiv, respiratory, strain, sars-cov-2\\
\hline
\end{tabular}
\caption{
Composition of $n=14$ Louvain clusters extracted from the undirected graph of the Faculty of Medicine \& Dentistry.
Department labels: a = Anatomy and Cell Biology, b = Biochemistry, e = Epidemiology and Biostatistics, mb = Medical Biophysics, mi = Microbiology and Immunology, ph = Physiology and Pharmacology, pl = Pathology and Laboratory Medicine.
}
\label{tab:faculty-clusters}
\end{table}

\end{document}